# The SONATE project, a French CANS for Materials Sciences Research

*Frédéric* Ott[1,*], *Alain* Menelle[1], *Christiane* Alba-Simionesco[1]

[1]Laboratoire Léon Brillouin CEA/CNRS, Université Paris Saclay, 91191 Gif sur Yvette Cedex, France

**Abstract.** We describe the Compact Accelerator-based Neutron Source SONATE which we are aiming for to replace the close Orphée reactor at Saclay, France. The SONATE source would serve an instrumental suite of about 10 instruments. The instruments would be split into low resolution instruments and higher resolution instruments. Our reference design is based on a proton accelerator operating at an energy in the range 20-30 MeV. The accelerator would serve 2 target stations. The first one operating at 20Hz with 2ms long pulses serving low resolution instruments (SANS, reflectivity, imaging, spin-echo) and the second one operating at 100Hz, 200µs long pulses serving higher resolution instruments (powder diffraction, Direct Time-of-flight spectroscopy, Indirect geometry spectroscopy). The 2 operation modes would be interlaced. The peak current on the target is aimed at 100 mA with an average power on the target on the order of 50-80 kW. Numerical Monte-Carlo simulations show that we may expect instrument performances equivalent to the current instruments around Orphée or ISIS.

## 1 Introduction

The reactor Orphée at the CEA Saclay has been serving the French neutron scattering community for 38 years. The reactor will however stop operation in 2019. The Laboratoire Léon Brillouin has been operating up to 24 neutron scattering instruments. The reactor was also serving other purposes such as industrial radiography, silicon doping and irradiation. We are considering replacing the reactor neutron source with a compact accelerator-based source which would serve an instrumental suite of about 10 instruments. In the following we describe the design parameters of a Compact Accelerator-based Neutron Source which would be suitable to replace the reactor Orphée for neutron scattering purposes. This design is referred to as “SONATE”.

## 2 The SONATE design parameters

We will consider the different degrees of freedom in the design of a Compact Accelerator-based Neutron Source and derive machine parameters we think are suitable for a CANS dedicated to neutron scattering which would provide sufficient performances to be operated as a user facility. The discussion is limited to neutron produced by the stripping reaction corresponding to proton energies below 50 MeV impinging a Lithium or Beryllium target.

### 2.1. Technological boundaries

#### *2.1.1 Accelerator technology*

Accelerators can operate in continuous mode or in pulsed mode. For neutron scattering the most efficient operation is in pulsed mode to benefit from the time-of-flight techniques. Most spallation sources are nowadays operating in pulsed mode which allows making use of most of the produced neutrons. In such pulses machines, the key figure of merit is the peak flux which is given by the proton peak current.

In the ESS design [1] the peak current was set to a conservative value of 62 mA. New facilities such as IFMIF/EVEDA have been operating at currents of up to 125 mA [2]. We will thus assume that it is possible to operate reliably an LINAC accelerator at a peak current of 100mA even though such high intensity accelerators are not common.

The choice of the ion particle is still under debate. The nuclear data are rather scarce in the 3-60MeV range and the cross sections are poorly known. Experimental work is under way to fill the database gap in the energy range of interest [3]. First results suggest that the neutron yield gains obtained with deuterons are no higher than 50% at best which is hardly enough to justify using deuterons over protons. Protons are significantly easier to handle than deutons and also they do not induce activation issues in the accelerator parts as deuterons do. Hence in the following we will only consider protons.

#### *2.1.2 Target material*

The choice of the target material is non-trivial and would require an extensive discussion. The situation can be roughly summarized as follows:

* Corresponding author: frederic.ott@cea.fr

- For proton energies below 3MeV, lithium should be preferred due to the low stripping reaction threshold. This is typically the choice made for low energy Boron Neutron Capture Therapy (BNCT) accelerators for example.
- For proton energies in the range 3MeV- 60MeV, beryllium and lithium are roughly equivalent in terms of neutron yield; However using significant power levels (kW) on the target will require to handle molten lithium which is challenging but is justified for very high power since it makes the cooling more efficient (see IFMIF@5MW). Beryllium on the other hand has a very high melting point (1287°C) and allows operating a solid target up to power densities in the range 0.5-1 kW/cm².
- From 60-100 MeV, there are several candidate materials (among which carbon for example). Again the lack of nuclear data makes choices non trivial.
- Above 100MeV, the neutron production via spallation channels starts being efficient and a heavy material target becomes the best choice (e.g. Tantalum).

A key choice in a facility design is thus the choice of the proton energy. This choice must be weighted by different factors: (i) the neutron yield, (ii) the accelerator cost, (iii) the power deposited on the target.

## 2.2 Choice of the proton energy

In the case of beryllium, there is a threshold of 2 MeV for neutron production and above 20 MeV the yield is roughly proportional to the proton energy: Yield ~ $E_p$ - 12.

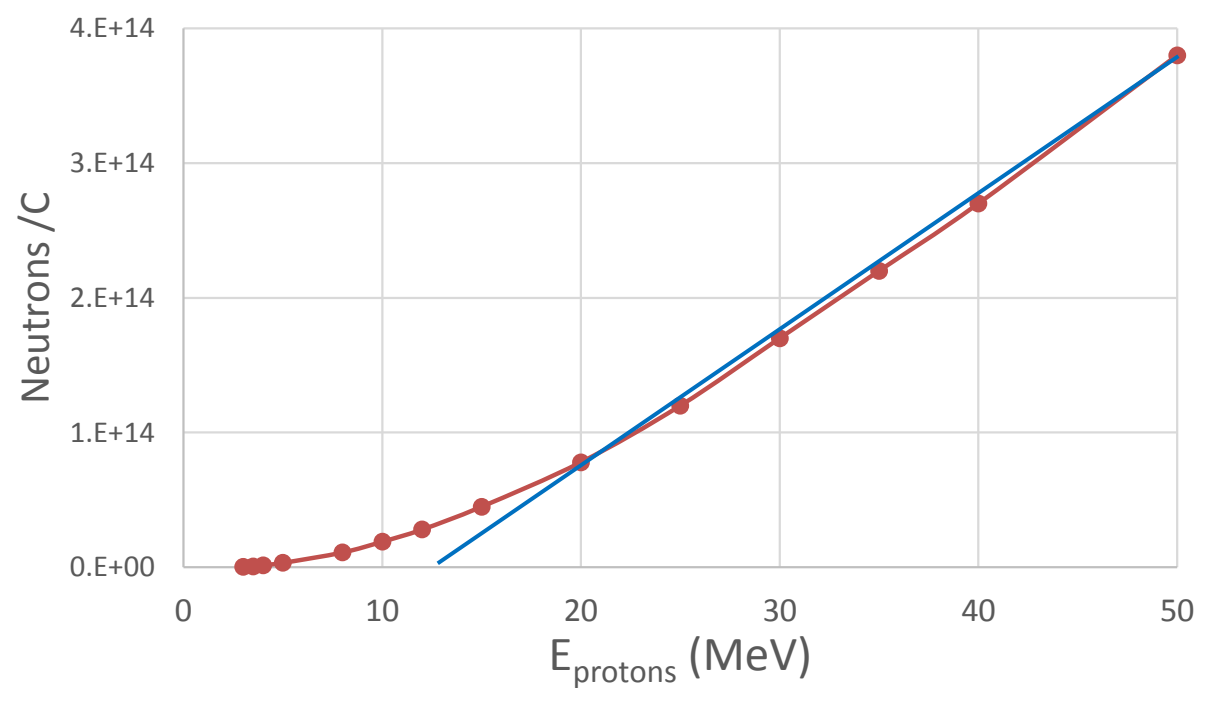

**Fig. 1.** Neutron yield as a function of the proton energy on a beryllium target (ENDF database).

The design of an efficient TMR requires that the target is as small as possible. A typical (maximum) size is on the order of 100 cm². Thermo-hydraulics requires that the maximum power density is in the range 0.5-1 kW/cm² which sets a limit of the ion beam energy at 50-100 kW. The Fig. 2 shows the power deposited on the target as a function of the proton energy. The orange line at P= 50 kW corresponds to a "safe" limit where thermo-hydraulics design are not too challenging. The limit at P=100 kW is probably a hard limit above which a fixed target cannot be used and a rotating target is necessary. The different load lines correspond to increasing duty cycle of the source.

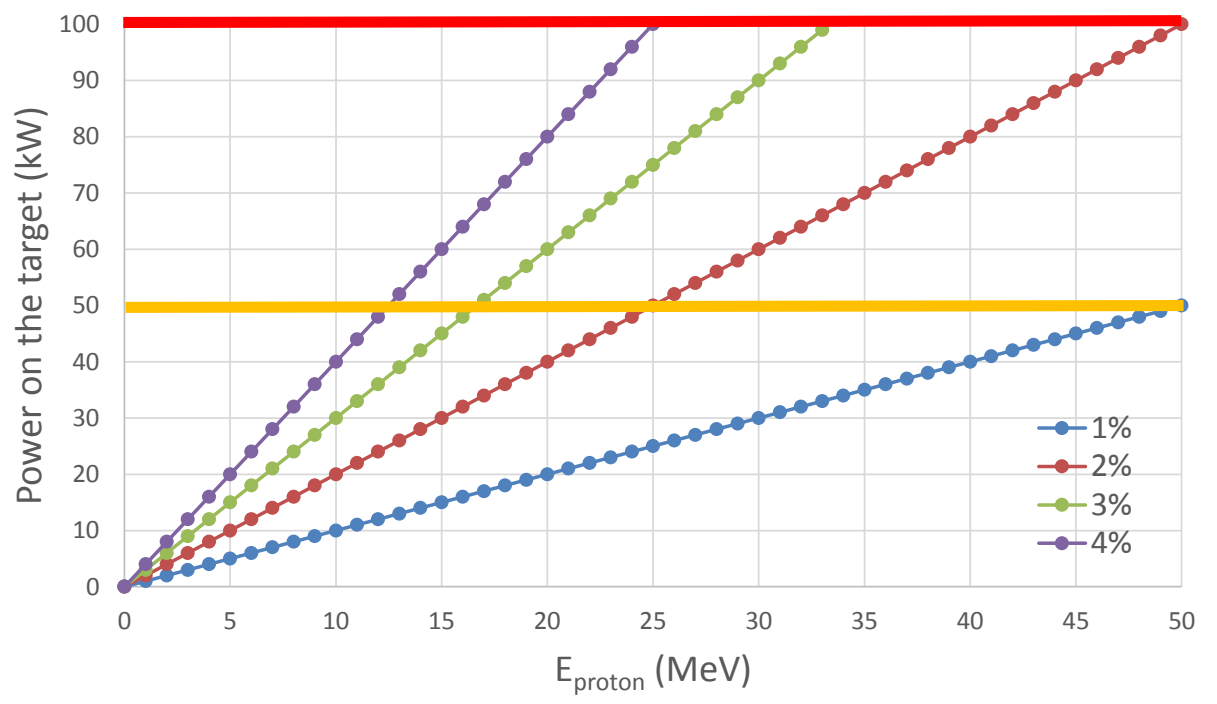

**Fig. 2.** Power on the target for increasing proton energy and various duty cycles from 1% to 4%.

As a reference, the long pulse source ESS is operating with 2.86ms pulses at 14 Hz corresponding to a duty cycle of 4%. In this operation scenario, the maximum proton energy which can be used is about 20-25 MeV. If the proton energy is increased to 50 MeV, the duty cycle has to be reduced below 2% otherwise the power density on the target would exceed reasonable levels. These figures are of course only indicative since the 100kW limit may be either difficult to achieve or may be overcome depending on the technology used; Nevertheless it illustrates that compromises have to be made between the proton energy and the duty cycle for a given power on the target. This is illustrated on Fig. 3a. For example, for a proton energy of $E_p$ = 20MeV, a peak current of 100 mA and a 4% duty cycle, the power on the target would be 80 kW and the neutron yield would be $3.1\times10^{14}$ n/s. If the proton energy is increased by a factor 2 to 40 MeV while the power on the target remain limited at 80kW, the duty cycle has to be reduced to 2% and the neutron yield is $5.4\times10^{14}$ n/s. For a given power deposited on the target, the neutron yield is roughly proportional the proton energy.

It is difficult to define a simple figure of merit since a number of "soft" parameters also play a role:

- What is the most suitable duty cycle? This depends on the aimed applications.
- For lower energy protons, the fast neutron spectrum is less energetic and thus easier to moderate.
- The gamma background is also less energetic for lower energy protons
- A lower energy accelerator is cheaper to build
- Above $E_p$ = 30MeV, new activation channels open and lead to activation of accelerators parts which can make the maintenance more complicated.

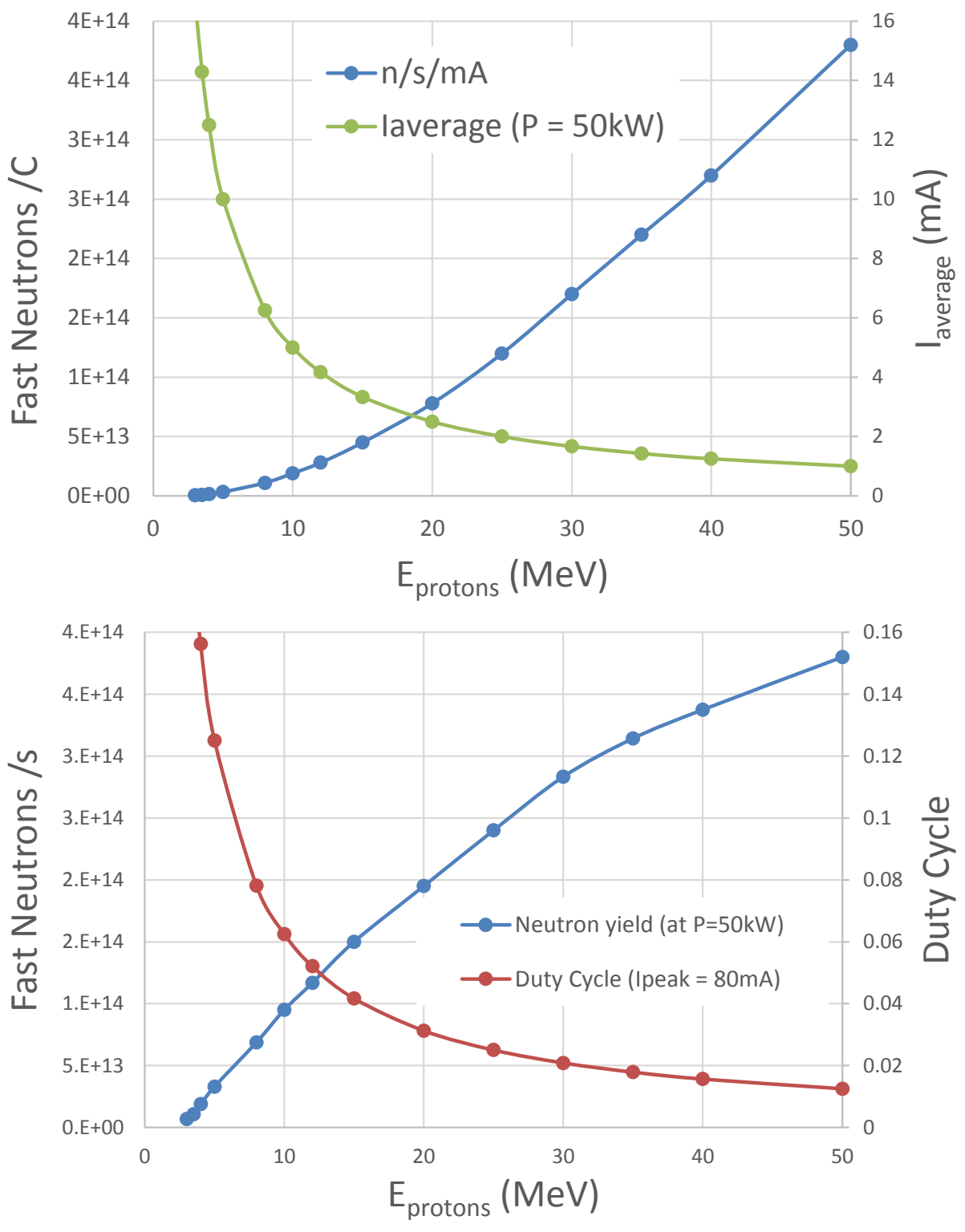

**Fig. 3.** (a) Neutron yield /C versus proton energy and maximum average current to limit the power on the target at 50kW. (b) Actual yield with a power limit on the target set at 50kW and corresponding duty cycle (for $I_{peak}$ = 80mA). While the neutron yield per proton doubles from $E_p$ = 20 to 30 MeV, power limits on the target reduce the gain to 45%. In parallel, the duty cycle on the target is reduced (from 3% to 2% if $I_{peak}$ = 80mA) which makes the neutron pulses easier to exploit for scattering experiments. From 20 MeV to 30 MeV the gain in flux is proportional to the accelerator cost. From 30 MeV to 50 MeV, the gain in flux is 36% while the proton energy is increased by 67%.

Assuming that the accelerator cost and operation is proportional to its energy (which is rather crude), a simple figure of merit could be defined as FOM = [Neutrons Yield/ $E_{protons}$] (see Fig. 4) which reflect the cost per produced neutron. It is clearly efficient to work above 10 MeV. However, even though the neutron yield per proton increases quickly with the proton energy, the figure of merit nevertheless decreases slowly above 20 MeV.

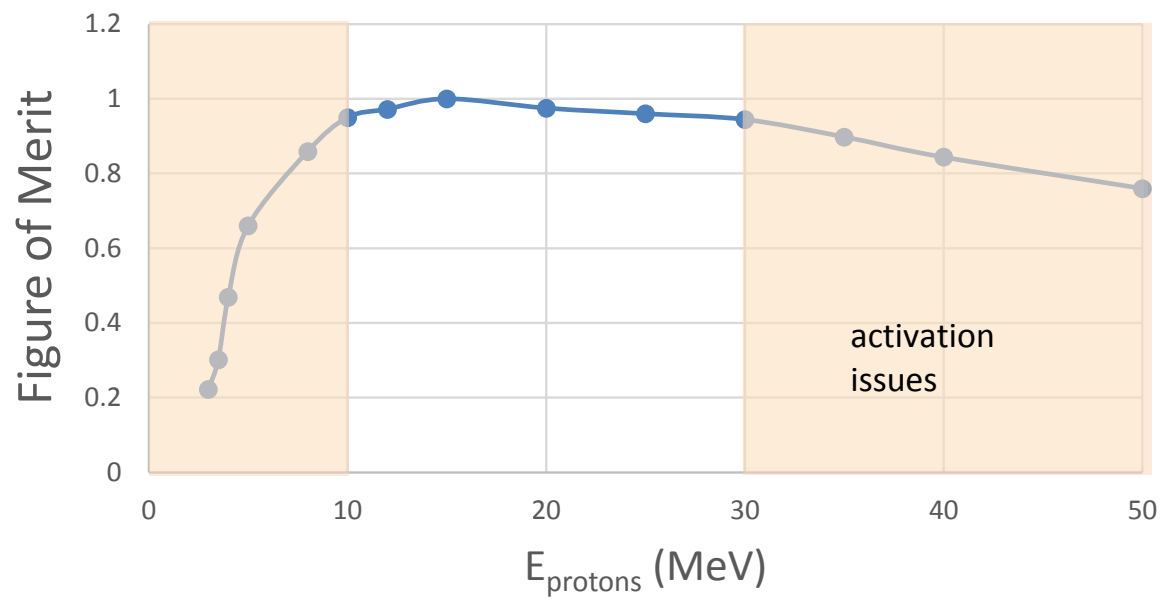

**Fig. 4.** Figure of merit defined as [Neutron Flux / Cost] (a.u.)

Below $E_p$ = 15 MeV, the maximum duty cycle is above 4% which becomes difficult to exploit efficiently for ToF neutron scattering. Above 30 MeV, proton activation channels open up and make the accelerator maintenance more complicated. Hence an optimal operation region for CANS seems to be in the range 10 – 30 MeV.

## 2.2 The SONATE design parameters

From the above considerations, we decided to set the SONATE reference parameters as:

- $E_p$ = 20MeV,
- $I_{peak}$ = 100mA,
- duty cycle = 4%, P = 80kW,
- fixed Be target.

These parameters were partly chosen because they correspond to the first 20m of the ESS Linac (out of 600m). Hence the components (Source, RFQ and DTL) are available with no R&D developments.
From a French nuclear regulation perspective, installations producing ionizing particles may be classified as “Installations Classées pour la Protection de l’Environnement” (ICPE) or “Installations Nucléaires de Base” (INB) [4]. In the latter case, fall of course research reactors (such as Orphée@Saclay) and particles accelerators (such as SPIRAL2@Caen) producing radionuclides above some legal thresholds. The INB categories are subject to very stringent rules which make their exploitation difficult.
Calculation of the activation products produced within the SONATE indicate that the facility would be considered as a simple ICPE.
In other countries, other rules and thresholds apply.

# 3 The expected performances for neutron scattering

In order to have an estimate of the performances a source such as SONATE could provide. A moderator design using polyethylene as moderating medium and beryllium as a reflector was considered. Moderation calculations were performed both with MCNP and GEANT4 [5]. In the case of the SONATE design parameters, a brilliance of $1.2x10^{11}$ n/cm²/s/sr was calculated and was used as an input in the instrument Monte-Carlo simulations (using McSTAS).
The neutron flux at the sample position for various neutron scattering techniques was calculated using simple ToF instrument designs [6] or even considering existing instruments around Orphée which were simply “moved” around SONATE. The results are summarized in the table below and compared with “reference” instruments at sources such as Orphée@Saclay or ISIS. The figures for the inelastic instruments (Direct TOF and Backscattering) have been taken from [7]. The orange figures correspond to various types of instrumental upgrades which are either in progress at the LLB or being implemented on some of the ESS instruments. In the case of reflectivity, the SELENE@ESS design could be implement to increase reflectivity measurement efficiency by an order of magnitude on small samples. In the case of SANS, focussing SANS

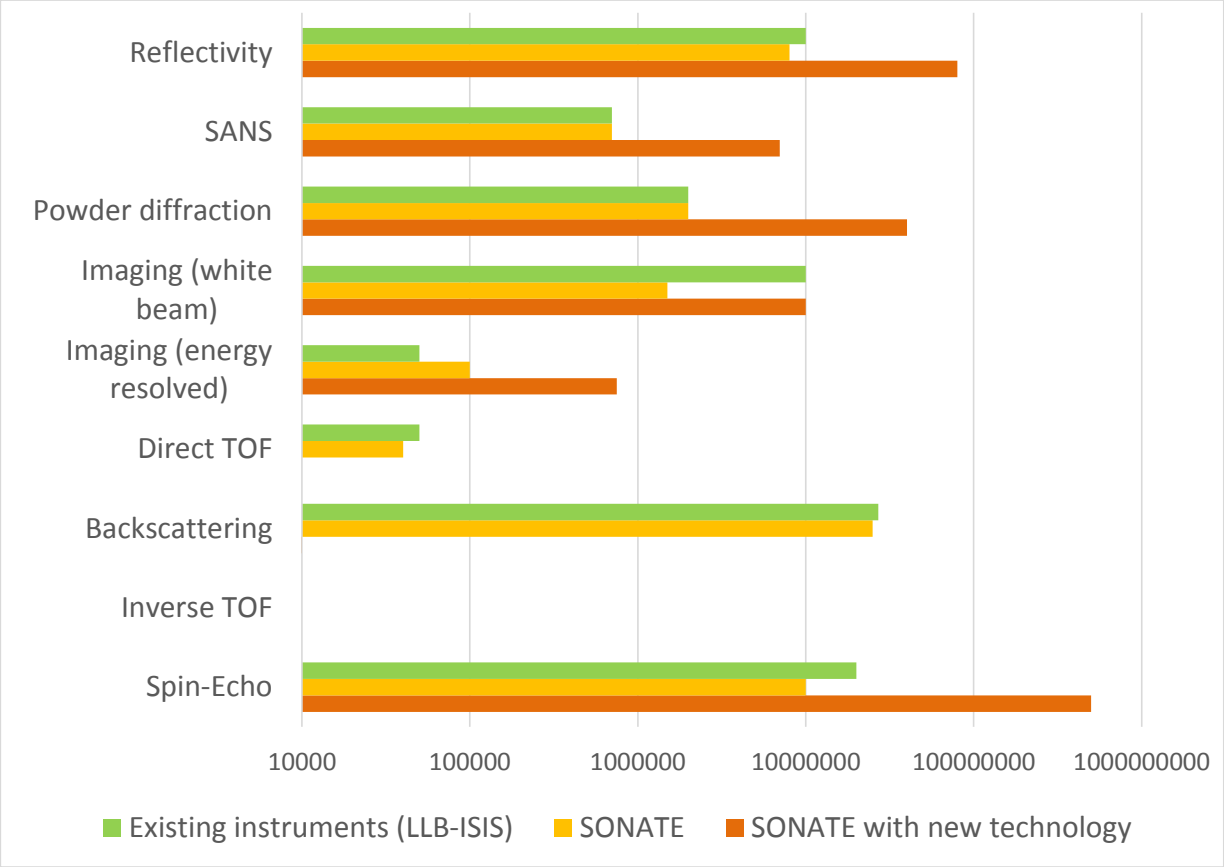

**Fig. 5.** Comparison of the performances of different scattering techniques in terms of flux at the sample position (n/cm²/s). (green) reference instruments at various facilities (LLB – ISIS) ; (yellow) move of the existing instruments from Orphée to SONATE ; (orange) performances after technical upgrades. The figures for the inelastic instruments (Direct TOF and Backscattering) have been taken from [7].

While it may be surprising that a CANS may achieve performances on par with existing medium scale research reactor or spallation sources, the reason for these performances can be qualitatively explained by the following reasons:

- The proton energy remaining low in a CANS, using a very high peak current is not detrimental to the overall (electrical) energy consumption of the source. While the ISIS TS2 is operated with an average current of 60µA, SONATE would be operated at an average current of 4mA. This large current compensates a large fraction of the neutron production efficiency difference between stripping and spallation.
- The moderator design is such that it is almost fully coupled to the fast neutron source. This is possible due to the small size of the primary fast neutron source, to the rather small heat load of the source but also from the rather low radiative gamma heating which is low is CANS especially compared to spallation. In CANS, the gamma spectrum is limited to the proton energy (a few 10 MeV) while in a spallation source high energy gammas are present in the source. The tight coupling between the source and the moderator leads to a gain of approximately a factor 5 compared to the coupling of the ESS moderator.
- The moderator design can make use of modern moderator design such as tube moderators as proposed recently [8]. A gain of a factor 5 relative to moderators at current sources is also expected.
- For each type of instrument, the source time structure has been optimized to fully fill the phase space, that is the pulse length and the repetition rates have been considered as free parameters and hence for each instruments almost all the neutrons can be used for scattering. Since the shielding constrains are very low, the chopper systems can be in principle set very close to the source (as close as 1m in theory) leading to additional gains compared to larger facilities.

Around existing sources, the time structure is usually fixed and better suited for specific instruments. For examples, short pulse spallation sources are well suited to high resolution experiments while short pulses are inefficient for low resolution experiments such as SANS, reflectometry or imaging. In the case of ESS, the operation parameters (2.6 ms, 14 Hz) were chosen as a weighted compromise to serve a very wide range of instruments. It may be argued that a CANS would face similar choices. The key difference is that the design and construction of a Target – Moderator – Reflector (TMR) assembly should have a rather limited cost (<1M€). It is even considered that the best option would be provide each instrument with its dedicated tube moderator [9]. A rather easy way to optimize the source time structure is to build several target stations with optimized instrumentation. The proton beam structure would have the time structure illustrated on Fig. 6 which interleaved long pulses at a low repetition rate with short pulses with a high repetition rate.

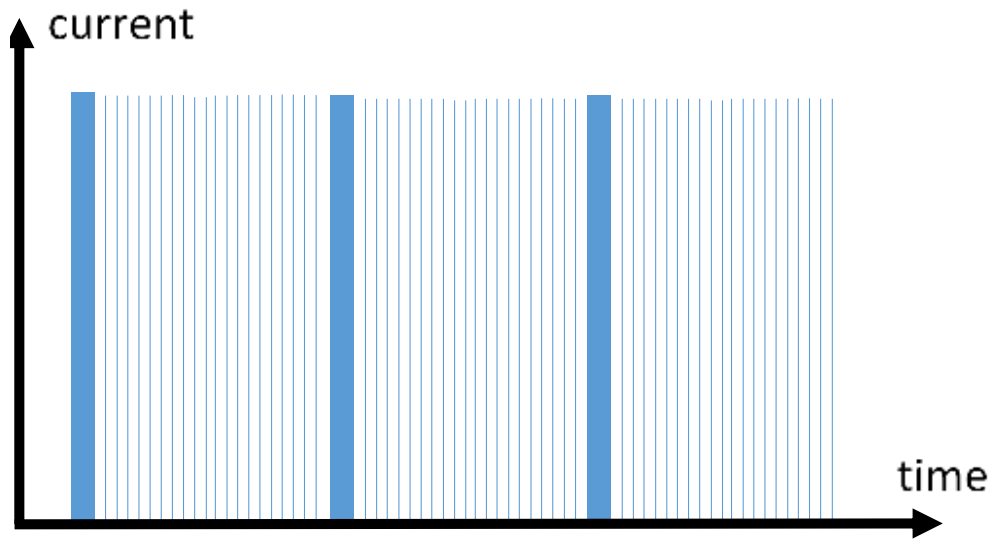

**Fig. 6.** Example of the proton pulse structure on SONATE. Long pulses (2ms, 20Hz) are interleaved with short pulse high repetition rate pulses (200µs, 100Hz). The long pulse are directed to a TMR station for low resolution instruments while the short pulses are directed to a second TMR station for higher resolution instruments. The first target would use 4% duty cycle while the second target would use 2% duty cycle. Hence the accelerator should be designed to handle a 6% duty cycle.

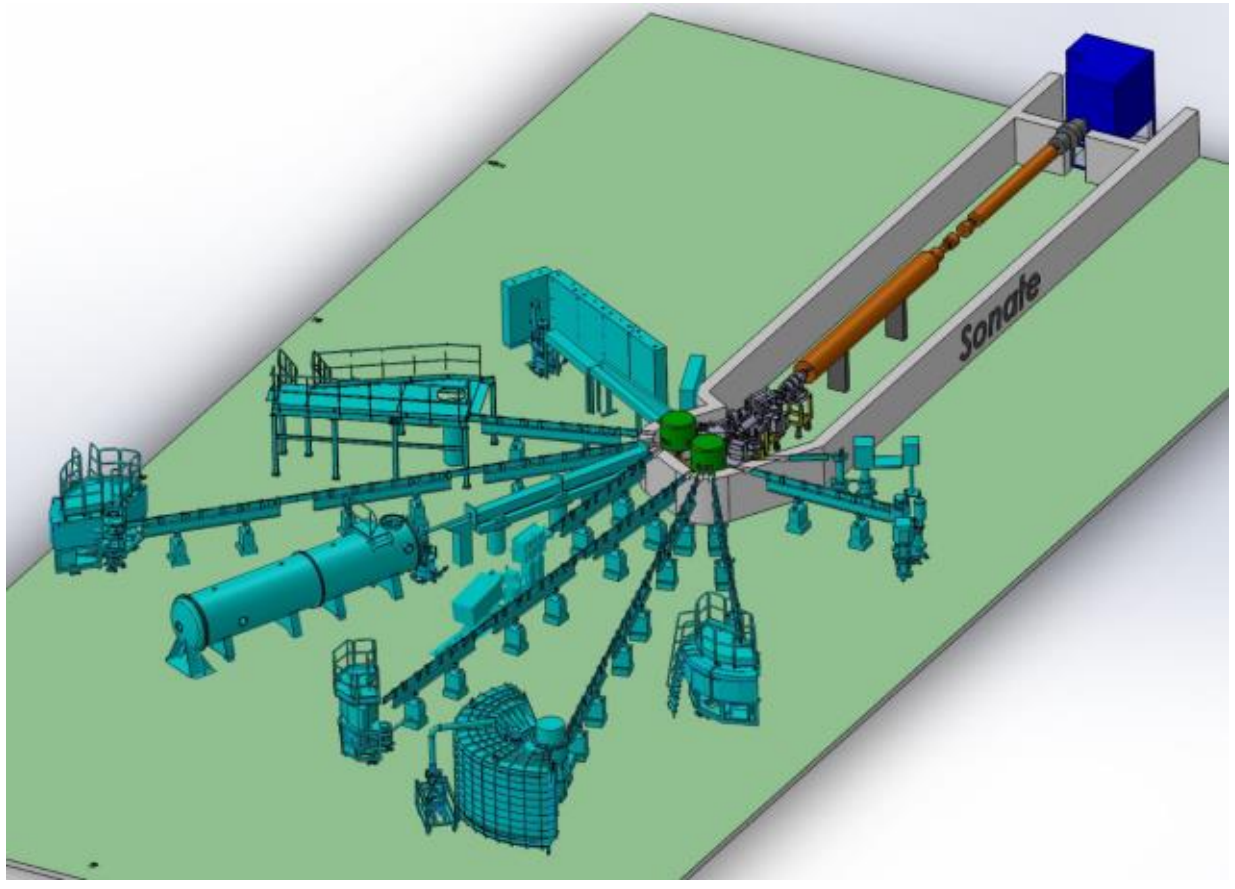

**Fig. 7.** A possible sketch of SONATE. A first long pulse / low repetition rate target station (green) would serve low resolution instruments such as an imaging station, a SANS, a reflectometer, a spin-echo and a low resolution powder diffractometer. A second target station with short pulses and high repetition rate would serve higher resolution instruments such a direct TOF instrument, a high resolution powder diffractometer, an inverse TOF diffractometer.

## 3 Conclusion

With the foreseen loss of neutron capacity in Europe due to the closure of aging neutron research reactors, alternative solutions must be found to continue providing neutron to neutron scattering users. We think that current accelerator technology is mature enough to build CANS which can provide neutron for scattering instruments and have performances on par with current state of the art medium scale research reactors or medium scale spallation sources. Beyond providing an alternative, these CANS represent an investment which is only a fraction of the cost of new nuclear reactors or spallation facilities together with reduced operation costs. Hence there is the possibility that the CANS technology may even allow an easier access to neutron scattering compared to the current situation. In which case a network of CANS across Europe could support an extended user community. This would be beneficial for the efficient use of the future European Spallation Source.